\documentclass[twocolumn]{aastex62}

\usepackage{multirow}
\usepackage{amsmath}

\def\iso#1{$^{#1}$}
\def\msun{$M_\odot$}

\submitjournal{ApJ}

\shorttitle{Origin of large meteoritic SiC}
\shortauthors{Lugaro et al.}

\begin{document}

\title{ORIGIN OF LARGE METEORITIC SIC STARDUST GRAINS IN METAL-RICH AGB STARS}

\correspondingauthor{Maria Lugaro}
\email{maria.lugaro@csfk.mta.hu}

\author{Maria Lugaro}
\affiliation{Konkoly Observatory, Research Centre for Astronomy and Earth Sciences, Konkoly Thege Mikl\'{o}s \'{u}t 15-17, H-1121 Budapest, Hungary}
\affiliation{ELTE E\"{o}tv\"{o}s Lor\'and University, Institute of Physics, Budapest 1117, P\'azm\'any P\'eter s\'et\'any 1/A, Hungary}
\affiliation{School of Physics and Astronomy, Monash University, VIC 3800, Australia}

\author{Borb\'ala Cseh}
\affiliation{Konkoly Observatory, Research Centre for Astronomy and Earth Sciences, Konkoly Thege Mikl\'{o}s \'{u}t 15-17, H-1121 Budapest, Hungary}

\author{Blanka Vil\'agos}
\affiliation{Konkoly Observatory, Research Centre for Astronomy and Earth Sciences, Konkoly Thege Mikl\'{o}s \'{u}t 15-17, H-1121 Budapest, Hungary}
\affiliation{ELTE E\"{o}tv\"{o}s Lor\'and University, Institute of Physics, Budapest 1117, P\'azm\'any P\'eter s\'et\'any 1/A, Hungary}

\author{Amanda I. Karakas}
\affiliation{School of Physics and Astronomy, Monash University, VIC 3800, Australia}
\affiliation{ARC Centre of Excellence for All Sky Astrophysics in 3 Dimensions (ASTRO 3D)}

\author{Paolo Ventura}
\affiliation{INAF - Osservatorio Astronomico di Roma, Via Frascati 33, 00040 Monte Porzio Catone (RM), Italy}

\author{Flavia Dell'Agli}
\affiliation{INAF - Osservatorio Astronomico di Roma, Via Frascati 33, 00040 Monte Porzio Catone (RM), Italy}

\author{Reto Trappitsch}
\affiliation{Nuclear and Chemical Sciences Division, Lawrence Livermore National Laboratory, Livermore, CA 94550, USA}
\affiliation{NuGrid Collaboration, http://www.NuGridstars.org}

\author{Melanie Hampel}
\affiliation{School of Physics and Astronomy, Monash University, VIC 3800, Australia}
\affiliation{ARC Centre of Excellence for All Sky Astrophysics in 3 Dimensions (ASTRO 3D)}

\author{Valentina D'Orazi}
\affiliation{INAF Osservatorio Astronomico di Padova, Vicolo dell’Osservatorio 5, 35122 Padova, Italy}

\author{Claudio B. Pereira}
\affiliation{Observatorio Nacional, Rua General Jos\'e Cristino, 77 Sao Cristovao, Rio de Janeiro, Brazil}

\author{Giuseppe Tagliente}
\affiliation{Istituto Nazionale di Fisica Nucleare (INFN), Bari, Italy}

\author{Gyula M. Szab\'o}
\affiliation{ELTE E\"otv\"os Lor\'and University, Gothard Astrophysical Observatory, Szent Imre h. u. 112, Szombathely, Hungary}
\affiliation{MTA-ELTE Exoplanet Research Group, 9700 Szombathely, Szent Imre h. u. 112, Hungary}
\affiliation{Konkoly Observatory, Research Centre for Astronomy and Earth Sciences, Konkoly Thege Mikl\'{o}s \'{u}t 15-17, H-1121 Budapest, Hungary}

\author{Marco Pignatari}
\affiliation{E. A. Milne Centre for Astrophysics, Department of Physics \& Mathematics, University of Hull, HU6 7RX, United Kingdom}
\affiliation{Konkoly Observatory, Research Centre for Astronomy and Earth Sciences, Konkoly Thege Mikl\'{o}s \'{u}t 15-17, H-1121 Budapest, Hungary}
\affiliation{NuGrid Collaboration, http://www.NuGridstars.org}
\affiliation{JINA-CEE, Michigan State University, East Lansing, MI, 48823, USA}

\author{Umberto Battino}
\affiliation{School of Physics and Astronomy, University of Edinburgh, EH9 3FD, UK}
\affiliation{NuGrid Collaboration, http://www.NuGridstars.org}

\author{Ashley Tattersall}
\affiliation{School of Physics and Astronomy, University of Edinburgh, EH9 3FD, UK}

\author{Mattias Ek}
\affiliation{Bristol Isotope Group, School of Earth Sciences, University of Bristol, Wills Memorial Building, Queen’s Road, Bristol BS8 1RJ, United Kingdom}
\affiliation{Institute for Geohemistry and Petrology, ETH Z\"urich, Clausiusstrasse 25, 8092 Z\"urich, Switzerland}

\author{Maria Sch\"onb\"achler}
\affiliation{Institute for Geohemistry and Petrology, ETH Z\"urich, Clausiusstrasse 25, 8092 Z\"urich, Switzerland}

\author{Josef Hron}
\affiliation{Institute for Astrophysics, University of Vienna, 1180 Vienna, T\"urkenschanzstrasse 17, Austria}

\author{Larry R. Nittler}
\affiliation{Department of Terrestrial Magnetism, Carnegie Institution for Science,
Washington, DC 20015, USA}


\begin{abstract}


Stardust grains that originated in ancient stars and supernovae are recovered from meteorites and carry the detailed composition of their astronomical sites of origin. We present evidence that the majority of large ($\mu$m-sized) meteoritic silicon carbide (SiC) grains formed in C-rich asymptotic giant branch (AGB) stars that were more metal-rich than the Sun. In the framework of the $slow$ neutron-captures (the $s$ process) that occur in AGB stars the lower-than-solar \iso{88}Sr/\iso{86}Sr isotopic ratios measured in the large SiC grains can only be accompanied by Ce/Y elemental ratios that are also lower than solar, and predominately observed in metal-rich barium stars -- the binary companions of AGB stars. 
Such an origin 
suggests that these large grains represent the material from high-metallicity AGB stars needed to explain the $s$-process nucleosynthesis variations observed in bulk meteorites \citep{ek19}. 
In the outflows of metal-rich, C-rich AGB stars 
SiC grains are predicted to be small ($\simeq$\,0.2\,$\mu$m-sized); large ($\simeq \mu$m-sized) SiC grains can grow if the number of dust seeds is two to three orders of magnitude lower than the standard value of 10$^{-13}$ times the number of H atoms. We therefore predict that with increasing metallicity the number of dust seeds might decrease, resulting in the production of larger SiC grains. 

\end{abstract}


\section{Introduction}
\label{sec:intro}

Stellar grains (``stardust'') are recovered from primitive meteorites, where they were incorporated 4.6 Gyr ago during the formation of the Solar System after interstellar travel from their stellar sites of origin -- mostly asymptotic giant branch (AGB) stars and core-collapse supernovae -- to the molecular cloud where the Sun formed \citep{zinner14,nittler16}. Since their discovery more than 30 years ago these grains have provided  powerful constraints for the modelling of nucleosynthesis in stars and supernovae. Recently, it has also become evident that the fingerprint of the almost pure nucleosynthetic signatures carried by these grains into different Solar System bodies can be exploited to understand the evolution of the protoplanetary disk and the formation of the planets \citep[see, e.g.,][]{dauphas04,poole17,burkhardt19,nanne19,stephan19,ek19}. 

Both O-rich (such as Al$_2$O$_3$, MgAl$_2$O$_4$, and silicates) and C-rich stardust has been recovered, among which the most widely and accurately investigated are silicon carbide (SiC) grains \citep{bernatowicz87}. This is due both to their sizes, which can be large enough (up to a few to tens of $\mu$m) for single grain analysis, and because it is easier to recover them compared to the other phases. On top of the main elements, Si and C, these grains contain enough other trace elements to provide very accurate snapshots of the composition of their parent stars: from N, Ne, Mg, S, Ca, Ti, Fe, and Ni, to rare elements heavier than Fe, such as Sr, Zr, Mo, Ru, Ba, Eu, W, and Hf \citep[see, e.g.,][]{lewis90,lewis94,hoppe94,savina04,avila12,avila13,liu14a,liu15,trappitsch18}. 

It was recognized early that the vast majority of these grains originate from a population of AGB stars of low mass, typically in the range 1.5 to 4 \msun. These stars reach the condition C$>$O required to condense SiC in their envelope thanks to the mixing (third dredge-up, TDU) episodes that carry to the stellar surface C produced by He burning in the deep He-rich intershell layer between the H- and the He-burning shells \citep[see][for a review]{karakas14dawes}. Evidence for this origin lies in the \iso{13}C/\iso{12}C ratio distribution of the grains, which is similar to that observed in C-rich AGB stars \citep{hoppe97a}, and the fact that
the grains are strongly enriched in \iso{22}Ne, another typical product of He burning \citep{gallino90}, and in the isotopes of elements heavier than Fe that are produced by the $slow$ neutron-capture process (the $s$ process) that occurs in the intershell of AGB stars \citep{lugaro03b}. However, the exact mass and metallicity range of the AGB parent stars of stardust SiC has been so far difficult to determine accurately. The vast majority ($\gtrsim$ 90\%) of the grains (the ``mainstream'' SiC grains) are generally believed to have originated in C-rich AGB stars of metallicity around solar, while an origin in AGB stars of lower metallicities has been attributed to the less abundant populations Y and Z ($\sim$1\% of all SiC each) \citep{hoppe97,amari01a,zinner06}, although there are some inconsistencies with such an explanation \citep[see, e.g.,][]{lewis13,liu19}. The remaining SiC grains mostly belong to population A+B, of unclear origin \citep{amari01c,liu17a,liu17b,liu18expHCCSN}, and to population X from core-collapse supernovae \citep[e.g.,][]{liu18b}. 

Here we focus on the mainstream SiC grains, which are by far the most abundant. While high-precision data are available for such grains, unambiguous identification of their exact AGB origin is hampered by the fact that various uncertainties related to the modelling of stellar physical processes can mimic variations in stellar mass and metallicity. This results in a degeneracy of the nucleosynthetic solutions that can be found to match the same observed composition. Stellar physics uncertainties include: the treatment of convective mixing in all phases of the evolution, stellar rotation, the rate of the mass loss from the stellar surface, and the mixing processes leading both to the dredge-up of carbon, \iso{22}Ne, and the $s$-process elements to the stellar surface, and to the production of a region rich in the main neutron source, \iso{13}C nuclei \citep[the \iso{13}C {\it pocket}, see discussion, e.g., in][] {cristallo09,piersanti13,trippella16,karakas16,liu18a,battino19,denhartogh19}. Therefore, in spite of their outstandingly high precision, isotopic measurements of SiC stardust cannot yet be used to strongly constrain uncertainties in stellar modelling (as we will discuss in Section~\ref{sec:Ni}). 

The best approach to identify the parent stars of the grains would be instead to compare the stardust data directly to the abundances derived from the spectra of AGB stars and their binary companions, for which we know the stellar mass and metallicity. Recently, a consistent spectroscopic data set for a large ($\simeq$180) sample of giant barium (Ba) stars, i.e., stars that accreted $s$-process elements from an AGB companion has become available \citep{decastro16} with an improved calculations of the uncertainties \citep{cseh18}. This new data set has already allowed us to determine, for example, that stellar metallicity is the main determinant of the distribution of the elements heavier than Fe in AGB stars and that other effects such as rotation should play a secondary role \citep{cseh18}; this is in agreement with asteroseismology observations \citep{denhartogh19}. Here, we aim to exploit the Ba star data to determine more accurately the origin of stardust SiC grains. 


\section{Observational data}
\label{sec:obs}

Analysis of stellar spectra mostly provides $s$-process constraints in terms of elemental abundances;  analysis of stardust grains instead mostly provides $s$-process constraints in terms of isotopic ratios. This is because 
elemental abundances in stardust SiC are predominantly controlled by the chemistry of dust formation rather than by nucleosynthesis. Therefore, we need a method to relate the elemental abundance ratios observed in $s$-process-enhanced stars to the isotopic ratio measured in SiC grains. To this aim we consider two main $s$-process observable ratios, one related to Ba stars (and inherited from their AGB companions) and one related to SiC stardust. The AGB parent stars of the SiC grains correspond to stars born between 5 and 10 Gyr ago, as obtained by summing (i) the age of the Sun of 4.6 Gyr, (ii) the interstellar lifetime of the grains \citep[between 0.3 and 3 Gyr,][]{heck20}, and (iii) the lifetime of a star of mass between 1.5 and 4 \msun, believed to become C-rich at solar metallicity, roughly between 0.2 and 3 Gyr. The Ba-star AGB companions probably represent a younger population since these stars are not required to have evolved to the AGB phase before the formation of the Sun. Their corresponding ages would have been roughly between 0.2 and 3 Gyr, plus the time elapsed since the mass transfer event that produced the Ba stars, which unknown. In any case, the main features (mass and metallicity) of the AGB parent stars of the SiC grains are comparable to those of the Ba-star AGB companions because we do not expect major variations in the initial mass function over the past 8 Gyr, nor do we observe major changes in the metallicity distribution, see, e.g., Figure 16 of \citet{casagrande11} and Figure~10 of \citet{hayden15}. 

\subsection{The [Ce/Y] ratio in Ba stars}
\label{sec:bastars}

\begin{figure}
\center{\includegraphics[width=1.05\linewidth]{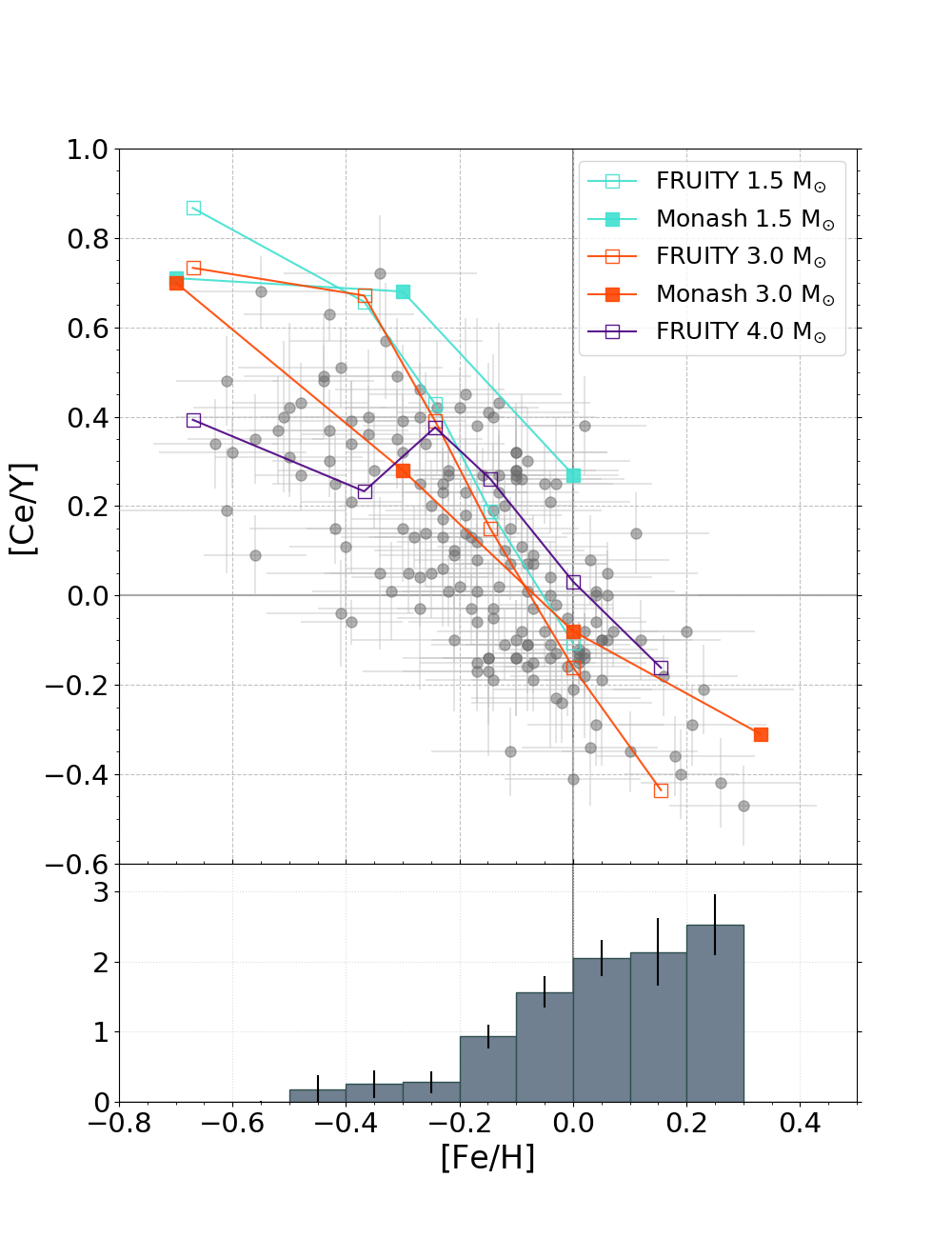}}
\caption{Top panel: the [Ce/Y] ratio observed in Ba stars as function of the metallicity [Fe/H] from \citet{pereira11} and \citet{decastro16}. 
The data points are semi-transparent, therefore, they appear darker in denser regions of the plot. 
The coloured lines show the results from a selection of the AGB models presented in Figure~\ref{fig:butterfly} (with the same symbols and colours) as examples of stellar model predictions \citep[see also][]{cseh18}. Bottom panel: the distribution of Ba stars with [Ce/Y]$<$0 representing the ratio of the number of stars with negative [Ce/Y] in each metallicity bin to the total number of stars with negative [Ce/Y], divided by the ratio of the number of stars in each bin to the total number of stars (i.e., $\frac{\mathrm{negative\,in\,bin}}{\mathrm{total\,negative}}/
\frac{\mathrm{total\,in\,bin}}{\mathrm{total}}$). The error bars on the bins are calculated by applying the bootstrap method 
(see Sec.~\ref{sec:translate} for details). \label{fig:Bastars}}
\end{figure}

For Ba stars, we consider the [Ce/Y]\footnote{Throughout the paper to express elemental ratios we use the square bracket notation, which represents the log$_{10}$ of the observed abundance ratios relative to the solar ratio, equal to zero by definition.} ratio derived from spectroscopic observations because Ce and Y are $s$-process elements belonging to the $s$-process second (isotopes with magic number of neutrons 82) and first (magic number of 50) peaks, respectively. This means that their relative abundances are a measure of the number of neutrons captured per Fe seed during the $s$ process. The [Ce/Y] ratios for the large sample of Ba stars from \citet{pereira11} and \citet{decastro16} are plotted against metallicity in the top panel of Figure~\ref{fig:Bastars}, with metallicities ranging from roughly 1/4 of solar ([Fe/H]=$-0.6$) to twice solar ([Fe/H]=$+0.3$). The [Ce/Y] errors were calculated for all of the sample stars  as described in \citet{cseh18}. Observationally, the main trend of the [Ce/Y] ratio in Ba stars is to decrease with increasing metallicity. This agrees with the main feature of low-mass AGB $s$-process models (some examples of which are shown in the top panel of Figure~\ref{fig:Bastars}), which is that the main neutron source \iso{13}C does not depend on the metallicity, while the main neutron absorber, Fe, obviously increases with increasing metallicity \citep{clayton88,gallino98}. This results in a decrease of the number of neutrons captured per Fe seed at higher metallicity, because the number of neutrons produced remains the same (as the number of \iso{13}C nuclei) but as Fe increases, for each Fe atom there are fewer neutrons available to capture. Overall, at higher metallicity the final atomic mass reached by the neutron-capture process is lower 
and it is more difficult to reach the second $s$-process peak. This leads to higher Sr, Y, Zr abundances as compared to Ba, La, and Ce. The same trend shown by [Ce/Y] in Figure~\ref{fig:Bastars} is also shown by other second-peak to first-peak elemental ratios. Here, we decided to consider [Ce/Y] because it shows the smallest uncertainty in the observational data \citep{cseh18}. 

\subsection{The $\delta$(\iso{88}Sr/\iso{86}Sr) ratio in stardust SiC grains}
\label{sec:size}

For SiC grains, high-precision isotopic data are available, which accurately reflect the composition of the gas from which the grains formed. Here, we chose to consider the \iso{88}Sr/\iso{86}Sr ratio and will express all isotopic ratios throughout the paper in form of $\delta$ values, i.e., permil variations with respect to the solar ratio, which is zero by definition in $\delta$ value. Positive/negative $\delta$ values indicate ratios higher/lower than solar. E.g.,  $\delta$(\iso{88}Sr/\iso{86}Sr)=+500 means that the measured \iso{88}Sr/\iso{86}Sr ratio is 50\% higher than the solar ratio, while $\delta$(\iso{88}Sr/\iso{86}Sr)=$-$500 means that the measured \iso{88}Sr/\iso{86}Sr ratio is 0.5 of the solar ratio. We selected the \iso{88}Sr/\iso{86}Sr ratio because \iso{88}Sr is located on the first $s$-process peak and $\delta$(\iso{88}Sr/\iso{86}Sr) can vary substantially from negative to positive values when increasing the number of neutron captures per Fe seed, similarly to the [Ce/Y] ratio. Also increasing the neutron density has the effect of increasing the value of $\delta$(\iso{88}Sr/\iso{86}Sr) because the branching points at \iso{85}Kr and \iso{86}Rb can open when the neutron density reaches above $5 \times 10^{8}$ and $10^{9}$ n cm$^{-3}$, respectively, therefore by-passing the production of \iso{86}Sr \citep[see, e.g.,][]{vanraai12,bisterzo15}\footnote{We note that the contribution of the weak $s$ process in massive stars to the Sr in the Solar System is too uncertain \citep[see, e.g., Figure~11 of][] {pignatari10} to independently constrain the $\delta$(\iso{88}Sr/\iso{86}Sr) from the AGB stars that synthesized the $s$-process Solar System abundances. In general, if the contribution to \iso{86}Sr from the weak $s$-process was significant, then the AGB stars that synthesized the $s$-process Solar System abundances must have produced on average a positive $\delta$(\iso{88}Sr/\iso{86}Sr) value to counterbalance the effect of the $s$-process in massive stars. If instead the contribution to \iso{86}Sr from massive stars is marginal, or \iso{88}Sr is also produced significantly \citep[which may be the case when considering rotation in massive stars, e.g.,][]{prantzos18}, then AGB stars must have produced on average a negative or null $\delta$(\iso{88}Sr/\iso{86}Sr) value.}. 

\begin{deluxetable*}{cccccc}
\tablewidth{0pc}
\tablecaption{Mass-weighted mean particle size and abundance in parts per million by mass (ppm) for the KJ SiC grains fractions extracted from the Murchison meteorite \citep{amari94} for which data are available for either $\delta$(\iso{29}Si/\iso{28}Si), $\delta$(\iso{88}Sr/\iso{86}Sr), and/or $\delta$(\iso{138}Ba/\iso{136}Ba). (Fractions KJF and KJH also exist but no such data are available.) All the data represent the composition measured by analysing millions of grains together (i.e., in bulk), except in the case of the large (KJG) grains for which data for single grain data are also available. The latter are indicated in italics and in parenthesis as reminders that they should not be quantitatively compared to the bulk data. \label{tab:size}}
\tablehead{\colhead{Fraction} & \colhead{Size$^{a}$ ($\mu$m)} & \colhead{Abundance (ppm)} & \colhead{$\delta$(\iso{29}Si/\iso{28}Si)$^{b}$} & \colhead{$\delta$(\iso{88}Sr/\iso{86}Sr)$^{c}$} & \colhead{$\delta$(\iso{138}Ba/\iso{136}Ba)$^{e}$}}
\startdata
KJA & 0.24 -- 0.65 & 0.25 & 22.2 $\pm$ 1.6 & \\
KJB & 0.32 -- 0.70 & 1.97 & 24.6 $\pm$ 1.3 & 20 $\pm$ 15 \\
KJC & 0.42 -- 1.02 & 1.11 & 29.0 $\pm$ 2.1 & 3 $\pm$ 14 & $-$319 $\pm$ 9 \\
KJD & 0.54 -- 1.23 & 1.21 & 27.0 $\pm$ 2.5 & $-$18 $\pm$ 18 & $-$321 $\pm$ 10 \\
KJE & 0.70 - 1.65 & 0.97 & 31.8 $\pm$ 3.0 & $-$40 $\pm$ 15 & $-$348 $\pm$ 10 \\
& & & 39.7 $\pm$ 3.6$^{d}$ & & \\
KJG & 2.1 -- 4.5 & 0.36 & 50.0 $\pm$ 5.6 & ({\it 0 to $-$200$^{f}$}) & ({\it $-$200 to $-$400$^{f}$}) \\
\enddata
$^{a}$Observed range containing 90\% of the mass (omitting top and bottom 5\%). 
$^{b}$\citet{amari00}, except where indicated otherwise .
$^{c}s$-process component of bulk data \citep{podosek04}. 
$^{d}$\citet{hoppe96}. 
$^{e}s$-process component of bulk data \citep{prombo93}.
$^{f}$Range where the single grain data concentrate \citep{liu18a}. 
\end{deluxetable*}

Since the 1990s, a large fraction of the meteoritic SiC grain data has been reported in the literature from experiments on stardust isolated by acid dissolution of the Murchison CM2 meteorite, and specifically from the K series extracted by \citet{amari94}. These authors used centrifugation to finely separate SiC grains of different size ranges; and named the different samples KJA to KJH, in order of increasing grain size, as indicated in Table~\ref{tab:size}. 
The first analyses of the isotopic composition of elements heavier than iron present in SiC in trace amounts were possible only on bulk samples, i.e., on large (millions of grains) collections of SiC of a given fraction (i.e., size). With the development of Resonant Ionisation Mass Spectrometry \citep[RIMS,][]{savina03,stephan16} it has become possible to analyse with statistical significance the composition of trace elements in single SiC grains. 
The RIMS high-precision data provide the composition of each AGB parent star of each single grain, rather than the average over a whole population of AGB parent stars as for the bulk analysis. However, only relatively large grains ($> 1 \mu$m) contain enough atoms to be analyzed individually via RIMS. 

Interestingly, it was noticed since the first analyses that the average isotopic composition of several elements in the SiC grains vary with the grain size. This indicates that nucleosynthetic and dust-growth processes somewhat correlate in AGB stars, possibly as function of mass and/or metallicity\footnote{One more complex case is that of the noble gas Kr, where variations in isotopic ratios as function of the grain size are also impacted on by the velocity of the wind that implanted such Kr atoms in pre-existing grains, producing different results between the slow AGB wind and the fast post-AGB wind \citep{lewis94,verchovsky04,lugaro17}.}. In the case of Sr,  $\delta$(\iso{88}Sr/\iso{86}Sr) varies from positive in the KJB and KJC size fractions, to negative in KJD and KJE (Table~\ref{tab:size}), with an overall variation of 6\% \citep{podosek04}. RIMS measurements of individual 1--3\,$\mu$m SiC grains are mostly located in the region between $-200$ and 0 \citep{liu15,liu18a}. 


For $\delta$(\iso{138}Ba/\iso{136}Ba), also a decreasing trend was found moving from KJC to KJE (Table~\ref{tab:size}). Because both $\delta$(\iso{88}Sr/\iso{86}Sr) and $\delta$(\iso{138}Ba/\iso{136}Ba) involve a nucleus with a magic number of neutrons, it was recognized early that the number of neutrons captured per Fe seed should play a role in such variations, with the parent stars of the larger grains somehow experiencing a lower number of neutrons captured per Fe seed \citep[see also discussion in][]{ott90,zinner91,gallino97}. With respect to the data from single KJG grains, $\delta$(\iso{88}Sr/\iso{86}Sr) shows more significant variations than Ba: in the large (KJE and KJG) grains it is mostly negative, while the bulk value on the small grains (KJB) is clearly positive. For $\delta$(\iso{138}Ba/\iso{136}Ba), instead, the large (KJG) grains cover the whole range observed in all the different smaller grain size fractions (KJC to KJE, see Table~\ref{tab:size}). 

We note that terrestrial contamination could have affected all of the SiC measurements and would have shifted the measured isotopic ratios toward the solar values, relative to the true compositions, and the amount of contamination could potentially depend on the grain size. It is thus possible on the one hand that the solar $\delta$(\iso{88}Sr/\iso{86}Sr) value reported for the KJC bulk measurement may be an artifact, and on the other hand that the true $\delta$(\iso{88}Sr/\iso{86}Sr) for bulk KJB may be more positive than reported.  The latter possibility would strengthen the conclusion that smaller and larger grains have positive and negative $\delta$(\iso{88}Sr/\iso{86}Sr) values, respectively.


As mentioned above, the neutron density can also have a role in increasing  $\delta$(\iso{88}Sr/\iso{86}Sr), while it does not affect $\delta$(\iso{138}Ba/\iso{136}Ba) since there are no branching points active in AGB stars that can modify the relative isotopic abundances of \iso{138}Ba and \iso{136}Ba. 
However, the effect of the neutron density on $\delta$(\iso{88}Sr/\iso{86}Sr) is predicted to be of second order relative to the effect of the number of neutron captures per Fe seed. Considering, for example, Figures~4 and 15 of \citet{lugaro03b}, where the effect of branching points was deliberately switched off, the variations in $\delta$(\iso{88}Sr/\iso{86}Sr) due to the variations in the number of neutron captures per Fe seed explored in that paper cover a range of 600 permil (roughly from -400 to -200), while the variations in $\delta$(\iso{138}Ba/\iso{136}Ba) due to the same variations in the number of neutron captures per Fe seed cover a much smaller range, of 250 permil (from roughly -350 to -100). Therefore,  $\delta$(\iso{88}Sr/\iso{86}Sr) is a better indicator of the number of neutron captures per Fe seed than $\delta$(\iso{138}Ba/\iso{136}Ba) in the metallicity range of solar to twice solar considered here and relevant for the mainstream SiC grains\footnote{At lower metallicities a significant effect of the higher neutron captures per Fe seed is predicted on $\delta$(\iso{138}Ba/\iso{136}Ba), which also increases to positive values as the neutron captures per Fe seed increases. For example, $\delta$(\iso{138}Ba/\iso{136}Ba)=577 for a 3 \msun\ model at of $Z=0.0028$ from \citet{karakas18} and $-24$ and 719 for 3 \msun\ models of $Z=0.008$ and $Z=0.003$, respectively, from the FRUITY database. This is because at such metallicities the range of neutron captures per Fe seed experienced by the AGB material corresponds to that where the second $s$-process peak elements are more significantly produced, while the range we consider in this paper corresponds to that where the first $s$-process peak elements are more significantly produced \citep[see, e.g., Figure~1 of][]{travaglio04}.}. 
Therefore hereafter we focus on the $\delta$(\iso{88}Sr/\iso{86}Sr) variations, which are the most significant. 

\section{Translating stardust $\delta$(\iso{88}Sr/\iso{86}Sr) ratios into Ba-star [Ce/Y] ratios}
\label{sec:translate}
\begin{figure*}
\center{\includegraphics[scale=1.3]{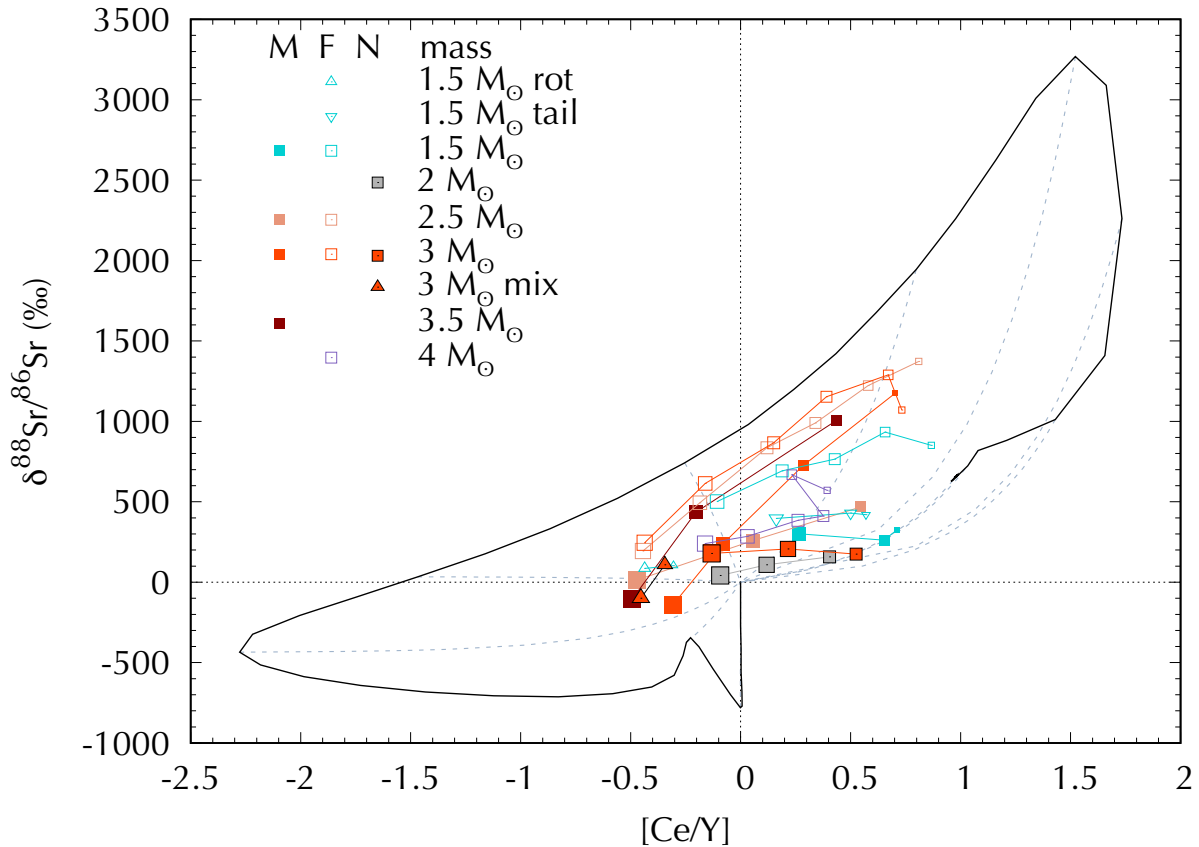}}
\caption{The $\delta$(\iso{88}Sr/\iso{86}Sr) as function of the [Ce/Y] ratio as predicted by a parametric model of neutron captures (black line). The initial abundances are scaled solar (i.e., the origin in the plot). Also shown by the colored symbols are predictions for the composition of the stellar surface at the end of the evolution from different sets of calculations of AGB stars of different initial masses (see legend). Same symbols of different sizes (connected by the solid, colored lines) indicate different initial metallicities (in the range listed below for each set of models), with increasing symbol size representing increasing metallicity. As discussed in Sec.~\ref{sec:bastars}, the general trend is that higher metallicity models typically predict lower [Ce/Y] ratios. We include models from the FRUITY database \citep[``F'', $Z=0.003-0.03$ and where ``rot'' refers to models rotating with initial velocity 60 km/s, and ``tail'' refers to models calculated with a different mixing profile leading to the formation of the \iso{13}C pocket,][]{cristallo15}, the Monash code \citep[``M'', $Z=0.0028-0.03$ and with the standard extent of the partial mixing zone leading to the formation of the \iso{13}C pocket as described in detail by][]{karakas16}, and the NuGrid collaboration \citep[``N'', $Z=0.01-0.03$ and where ``mix'' represent the case where a constant slow mixing is included inside the \iso{13}C pocket][]{battino19}. Note that several other models were presented by \citet{battino19} to simulate potential mixing, but we do not include them in the figure for sake of clarity as they overlap with the plotted models. Example of mixing lines between a selection of points from the parametric model and the solar composition are shown as dashed, grey lines. All the stellar model results are within such mixing lines.
\label{fig:butterfly}
}
\end{figure*}

\subsection{Parametric models (black solid line in Figure~\ref{fig:butterfly})}
\label{sec:parametric}

To compare the SiC to the Ba-star data and identify the parent stars of the grains, we need to investigate the correlation between  $\delta$(\iso{88}Sr/\iso{86}Sr) and [Ce/Y] predicted by models of the $s$ process. The more general the model is, the more robust the prediction of such correlation, therefore, we start considering a parametric model of the $s$ process, which is not affected by all the stellar model uncertainties mentioned in the Introduction. In Figure~\ref{fig:butterfly} we plot the results of a parametric model of $s$-process nucleosynthesis calculated by feeding a constant neutron density to a nuclear network with initial low-metallicity abundance distribution using the NucNetTools \citep{meyer12} and the JINA Reaclib database \citep{cyburt10}. \citep[More information and details on these models can be found in][]{hampel16,hampel19}. 

We set the constant neutron density to $10^7$ n cm$^{-3}$ and run the simulations for 60,000 yr, which are typical orders of magnitude for the neutron flux in the \iso{13}C pocket. If we increase the value of the neutron density, the branching points at \iso{85}Kr and \iso{86}Rb can activate and $\delta$(\iso{88}Sr/\iso{86}Sr) may increase. For example, for a neutron density of $10^8$ n cm$^{-3}$ the value of  $\delta$(\iso{88}Sr/\iso{86}Sr) in the section of the black line located in top right quadrant of Figure~\ref{fig:butterfly} increases and reaches $\simeq$ 12,000. 
In relation to this, it should be considered that in models of AGB stars with mass typically higher than 3 \msun, the activation of the \iso{22}Ne($\alpha$,n)\iso{25}Mg reaction during recurrent episodes of He burning may produce a further neutron burst with higher neutron densities than those within the \iso{13}C pocket, potentially activating the branching points at \iso{85}Kr and \iso{86}Rb and increasing the value of $\delta$(\iso{88}Sr/\iso{86}Sr). This neutron burst provides a small amount of neutrons and it only contributes to the production of the first peak $s$-process elements, at least in AGB stars of metallicity around solar, potentially marginally decreasing the [Ce/Y] ratio. Overall, an upward shift of 300 to 400 in $\delta$(\iso{88}Sr/\iso{86}Sr) \citep[compare Figures 3 and 4 of][]{lugaro03b} and a decrease in [Ce/Y] by up to -0.1 dex \citep[see discussion in][]{cseh18} would qualitatively account for the upper limit of the impact of the \iso{22}Ne neutron burst on each of the points on the black line shown in Figure~\ref{fig:butterfly}. 

We set a temperature of $0.9 \times 10^8$ K and a density of 1000 g/cm$^3$, typical for the \iso{13}C pocket. These specific choices do not have a significant effect on the final results since we are not modelling the production of neutrons, and because  
neutron-capture reaction rates are typically only  mildly dependent on the temperature, as the cross section is roughly proportional to the inverse of the velocity. 
We tested the impact of changing the temperature to $1.5 \times 10^8$ K. This has a minor effect and only on the maximum $\delta$(\iso{88}Sr/\iso{86}Sr), which decreases from 3300 to roughly 2500.


Overall, the results from such a parametric model are controlled almost exclusively by the general properties of neutron-capture cross sections of the isotopes on the $s$-process path, specifically the presence of isotopes with magic numbers of neutrons. As time passes, more and more neutrons are fed into the network and elements heavier than Fe are produced. Up to roughly 6000 yr (a neutron exposure of roughly 0.23 mbarn$^{-1}$), the first $s$-process peak elements represented by Y accumulate, since their magic number of neutrons results in relatively small neutron-capture cross sections ($\simeq$ a few mbarn), and [Ce/Y] is negative (bottom left quadrant of Figure~\ref{fig:butterfly}). After this time, the abundance of \iso{89}Y (the only stable isotope of Y) becomes large enough that this nucleus also starts capturing neutrons (top left quadrant of Figure~\ref{fig:butterfly}), producing the second-peak elements (e.g., Ce). Eventually, [Ce/Y] becomes positive (top right quadrant of Figure~\ref{fig:butterfly}). When Ce also becomes abundant enough to start capturing neutrons, the flux reaches the third peak at Pb (magic number of neutrons 126) and [Ce/Y] settles on an equilibrium value of roughly +0.8 dex. 

At the same time $\delta$(\iso{88}Sr/\iso{86}Sr) also evolves: it remains negative together with the [Ce/Y] ratio (bottom left quadrant of Figure~\ref{fig:butterfly}) but becomes positive before the [Ce/Y] ratio does (top left quadrant of Figure~\ref{fig:butterfly}). When the [Ce/Y] ratio reaches zero, $\delta$(\iso{88}Sr/\iso{86}Sr) is already $\sim +1000$. It reaches a maximum of roughly $+3300$ and then turns down to settle into the equilibrium value of roughly $+600$ (top left quadrant of Figure~\ref{fig:butterfly}). A main result is that during the $s$ process negative $\delta$(\iso{88}Sr/\iso{86}Sr) values are always accompanied by negative [Ce/Y], i.e., there are no model predictions in the bottom right quadrant of Figure~\ref{fig:butterfly}. This is simply because the second $s$-process peak at Ce is populated only after the first $s$-process peak at Sr, Y, and Zr. In other words, before the flux can proceed to the second peak, \iso{88}Sr needs to be overproduced relatively to \iso{86}Sr, relative to solar. The potential activation of the \iso{85}Kr and \iso{86}Rb branching points mentioned above does not change this overall conclusion since their only possible effect on the result of Figure~\ref{fig:butterfly} would be to increase the value of $\delta$(\iso{88}Sr/\iso{86}Sr).



\subsection{Stellar models (symbols connected by colored solid lines in Figure~\ref{fig:butterfly})}
\label{sec:stellar}

While simple parametric neutron-capture models as presented in the previous section do not produce realistic predictions for the surface composition of an AGB star, they still provide limits within which $s$-process stellar model predictions must be located. More realistic stellar models take into account several effects. First, the material in the He-rich intershell where the $s$ process occurs is not the result of a single episode of neutron captures, but of the combination of many cycles of neutron captures. In fact, a \iso{13}C pocket forms as a consequence of each TDU episode, of which there are typically 10-20 for C-rich stars in the low-mass AGB range considered here. Second, as mentioned at the start of the previous section, a marginal neutron flux can also occur within the recurrent convective instabilities during the episodes of He burning in the intershell due to the activation of the \iso{22}Ne reaction, and this can also affect the final intershell composition, particularly for the higher range of the stellar masses when considering the effect of the neutron density and the related operation of the branching points at \iso{85}Kr and \iso{86}Rb on the $\delta$(\iso{88}Sr/\iso{86}Sr) discussed qualitatively in the previous section.
Third, the material from the intershell is recursively carried to the stellar surface by the TDU, and thus is diluted with the envelope material. 

Examples of predicted final surface compositions from three different sets of AGB stellar models are shown in Figure~\ref{fig:butterfly}. As expected, the results from the stellar models lie on the intersection of mixing lines connecting the envelope abundances (with a distribution assumed to be close to solar) to compositions produced by different amount of neutrons within the parametric models. In other words, the AGB $s$-process models are constrained to lie within the area covered by ``butterfly'' shape produced by the parametric neutron-capture model and its mixing lines with solar composition. Only a few of the plotted stellar models reach the region of negative $\delta$(\iso{88}Sr/\iso{88}Sr) observed in the large SiC grains (bottom left quadrant of Figure~\ref{fig:butterfly}): the 3 and 3.5 \msun\ Monash models with $Z=0.03$ and the 3 \msun\ NuGrid models of $Z=0.03$ that include slow mixing in the \iso{13}C pocket. The several more models with negative $\delta$(\iso{88}Sr/\iso{88}Sr) presented by \citet{liu18a} support the results of Figure~\ref{fig:butterfly} (Nan Liu, personal communication).

Overall, the evidence from Figure~\ref{fig:butterfly} is that the negative $\delta$(\iso{88}Sr/\iso{86}Sr) values observed in the large SiC grains are necessarily accompanied in their parent stars by also negative [Ce/Y]. Therefore, to identify the parent stars of such grains we need to search for the AGB stars that correspond to the companions of Ba stars with [Ce/Y] lower than zero. Actually, this is a very conservative limit because the condition that [Ce/Y] is lower than zero is necessary but not sufficient, given that it is possible to find neutron-capture results that show [Ce/Y] lower than zero but $\delta$(\iso{88}Sr/\iso{86}Sr) ratios higher than zero (top left quadrant of Figure~\ref{fig:butterfly}).

\subsection{Comparison to the Ba star sample}
\label{sec:comparison}

In the bottom panel of Figure~\ref{fig:Bastars} we show the normalized distribution of the number of Ba stars that show negative [Ce/Y] for different metallicity bins. Each bin corresponds to 0.1 dex in metallicity and the error bars on each bin were calculated using the bootstrap method \citep{efron1979} as follows. We simulated 10,000 samples, with 182 randomly chosen stars for each run by applying ``random sampling with replacement'' on the whole sample data of 182 stars. This means that in each simulated sample some stars can appear more than once, while others can be missing. For each selected star, the [Ce/Y] ratio was chosen randomly from a normal distribution with a width corresponding to the error bar of its given [Ce/Y]. Finally, we calculated the number of stars in each metallicity bin for all runs and the final error on the height of a bin as the standard deviation of the 10,000 runs.

The distribution of Ba stars with sub-solar [Ce/Y] ratios, i.e., the  candidate parent stars of the large SiC grains, is heavily skewed towards stars of higher-than-solar metallicity, whose companions therefore appear to be the favoured site of formation of the grains with negative $\delta$(\iso{88}Sr/\iso{86}Sr). For example, stars of metallicity from 1.6 to 2 times solar, i.e., [Fe/H]=$+0.2$ to $+0.3$, or Z$\simeq$0.02 to 0.03 \citep[using the solar metallicity of 0.014 from][]{asplund09} are roughly 70\% more likely to be the parent stars of the large grains than are stars from solar to 25\% lower than solar metallicity (i.e., [Fe/H]=$-0.1$ to $0$, or Z$\simeq$0.01 to 0.014); and roughly 2.5 times more likely than stars with metallicity between 60 and 80\% of solar (i.e., [Fe/H]=$-0.2$ to $-0.1$, or Z$\simeq$0.009 to 0.01). This estimate is a lower limit because as the number of neutrons captured by Fe seed increases,  $\delta$(\iso{88}Sr/\iso{86}Sr) becomes positive before the [Ce/Y] ratio does (top left quadrant of Figure~\ref{fig:butterfly}). Because both the observational and theoretical [Ce/Y] versus [Fe/H] trends show that the number of neutrons captured per Fe seed increases as the metallicity decreases, it is more likely that stars of lower rather than higher metallicity may have a negative [Ce/Y] accompanied by a positive $\delta$(\iso{88}Sr/\iso{86}Sr).

\subsection{Discussion and further predictions}
\label{sec:discussion}

In summary, it appears that the larger SiC grains should form in AGB stars of higher metallicity than the AGB sources of the smaller SiC grains. Barium isotopic data from even larger SiC grains \citep[7-58 $\mu$m from the LS + LU fractions of][]{amari94} showing no Ba nucleosynthetic variation \citep{avila13} would then follow this trend and originate in AGB stars of even higher metallicity, such that the Ba isotopes remain mostly unaffected by $s$-process nucleosynthesis, however, they might still show variations in the composition of $s$-process elements belonging or close to the first $s$-process peak, e.g., Sr, Zr, and Mo. 

The hypothesis that larger grains should come from more metal-rich AGB stars also predicts that the Si isotopic ratios should present some variations with the grain size not due to nucleosynthetic processes, as for Sr and Ba, but due to galactic chemical evolution models, which predict that $\delta$(\iso{29}Si/\iso{28}Si) increases with the metallicity \citep{timmes96,kobayashi11a,lewis13}. Even though heterogeneities in the interstellar medium could somewhat smear out such an increase \citep{lugaro99,nittler05}, the expected trend is observed in the bulk data of SiC in grains of different sizes. These measurements show in fact that $\delta$(\iso{29}Si/\iso{28}Si) increases with the grain size (Table~\ref{tab:size}). Among the Ti isotopic ratios, the least affected by neutron captures in AGB stars is $\delta$(\iso{47}Ti/\iso{48}Ti), therefore, this ratio should also carry the signature of the initial composition of the star. While an increase with grain size is observed from the KJA/KJB grains towards the KJF grains, KJG and KJH grains show a decline, although contamination problems could have affected the data \citep{amari00}. 

In relation to the lighter elements C and N, interesting isotopic trends with grain size are present \citep{hoppe94,hoppe96}, however, they need to be discussed in a separate work because the isotopic compositions of these elements are affected by mixing processes in red giant and AGB stars, some of which are still not well understood \citep[see, e.g.][]{karakas14dawes}, and N may in some cases be severely affected by terrestrial contamination. The presence of the radioactive nucleus \iso{26}Al in SiC \citep{groopman15} also needs to be analysed in the light of the different grain sizes together with its production in AGB models of different metallicity. Both the Monash and the FRUITY models predict increasing \iso{26}Al/\iso{27}Al ratios with increasing stellar metallicity \citep{karakas16,cristallo15}. 

Finally, disentangling possible trends of metallicity versus grain size from observations of elemental abundances in SiC is more difficult than from isotopic ratios because, as mentioned at the start of Section~\ref{sec:obs}, elemental abundances are also affected by the chemistry of dust condensation around AGB stars of different metallicity \citep[see][and Sec.~\ref{sec:dust}]{amari95b}. In any case, available measurements may be in agreement with the metallicity versus size trend proposed here because (i) rare earth elemental abundances measured in bulk KJB grains are in agreement with models of metallicity around solar \citep{ireland18}, (ii) larger (KJH) grains contain in general lower elemental abundances than smaller grains \citep{amari95b,amari00}, and (iii) KJH grains with the highest $\delta$(\iso{29}Si/\iso{28}Si) ratios have also the lowest abundances of Ce relatively to Y \citep[see Figure~3 of][]{amari95b}.

\section{Comparison between AGB nucleosynthesis models and Ni and Ti grain data}

\label{sec:Ni}

\begin{figure*}
\center{\includegraphics[scale=0.7]{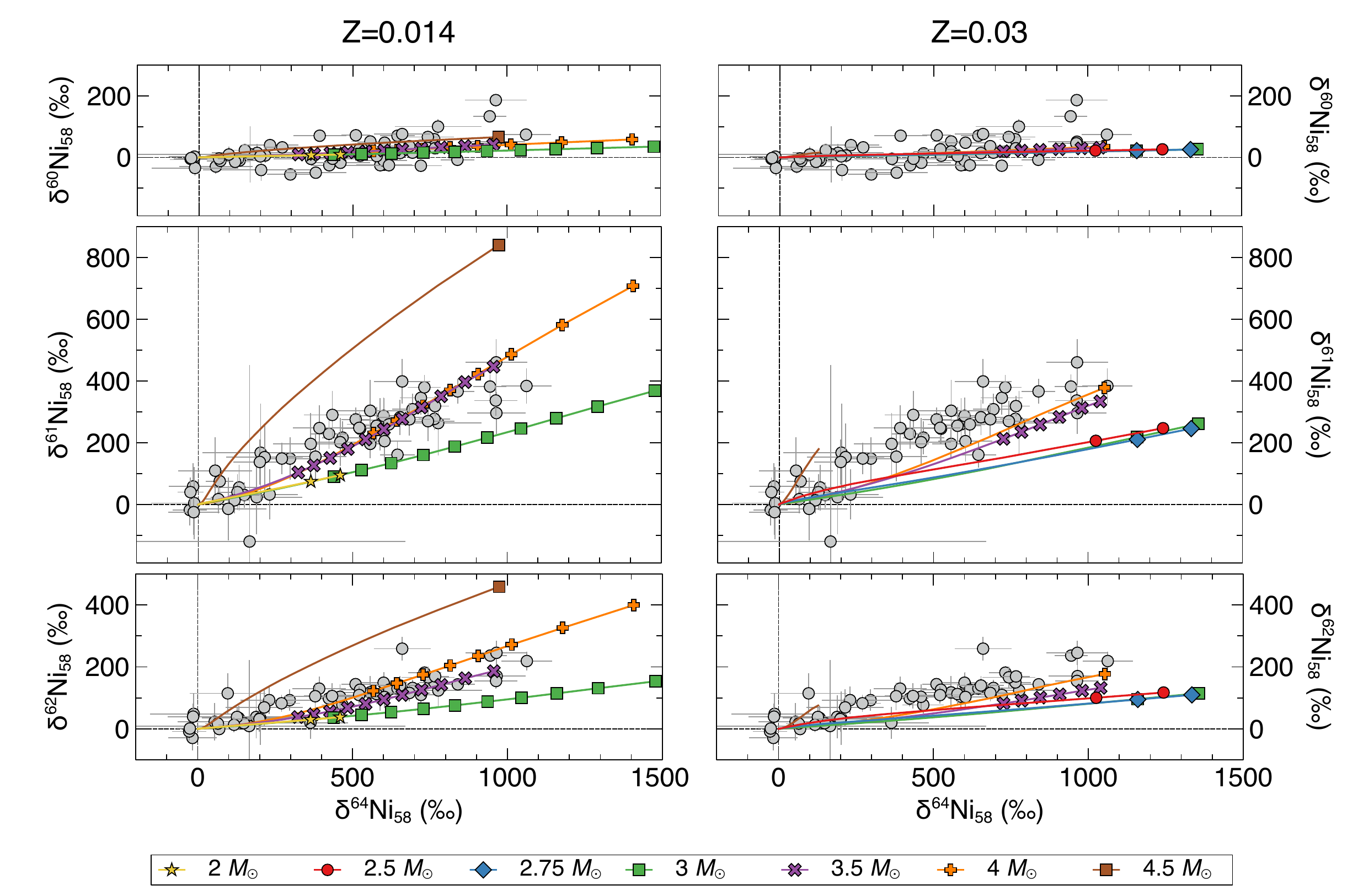}}
\caption{Comparison between single KJG SiC data from the Ni measurements of \citet{trappitsch18} (grey symbols with 2$\sigma$ error bars) and AGB stellar surface predictions from the models of \citet{lugaro18} at solar ($Z=0.014$, left panels) and twice solar ($Z=0.03$, right panel) metallicity and initial stellar masses between 2 and 4.5 \msun. Different masses are represented by different colours, and symbols are plotted on top of the lines only when C$>$O, the condition of SiC formation, is achieved in the envelope. Note that in axis labels of this and the following comparison figures a shortened version of the $\delta$ notation is used, e.g., $\delta^{60}$Ni$_{58}$ represents $\delta$(\iso{60}Ni/\iso{58}Ni). \label{fig:Ni_Z}}
\end{figure*}

\begin{figure}
\center{\includegraphics[scale=0.7]{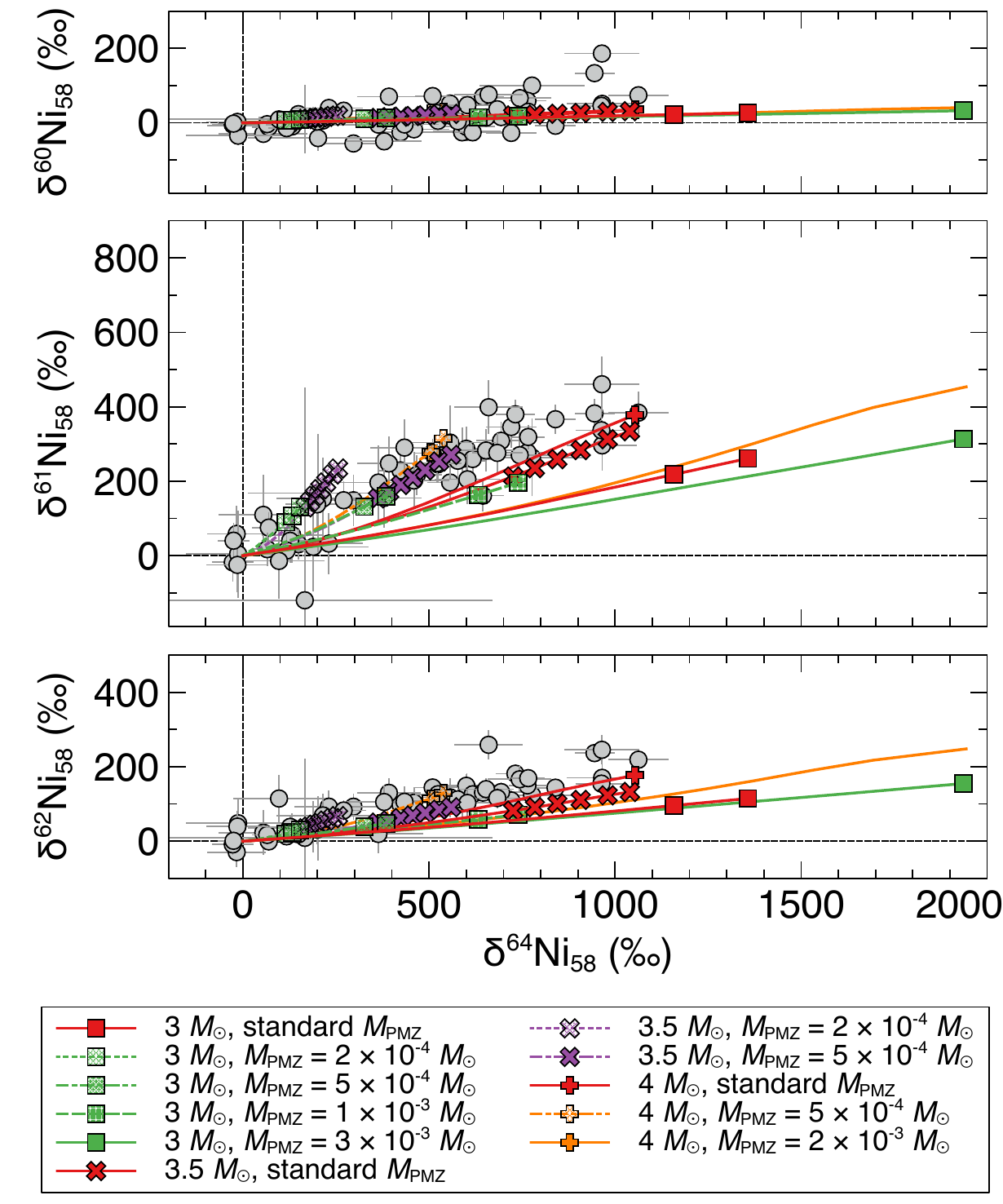}}
\caption{Comparison between single KJG SiC data for the Ni isotopes from \citet{trappitsch18} and the AGB model predictions of \citet{lugaro18} at twice solar ($Z=0.03$) metallicity, initial stellar masses between 3 and 4 \msun, and different extent of the mass of the mixing of protons required to form the \iso{13}C pocket ($M_{\rm PMZ}$). As in Figure~\ref{fig:Ni_Z} isotopic ratios are plotted on top of the lines only when the C$>$O is achieved in the envelope. \label{fig:Ni_PMZ}}
\end{figure}

In Figure~\ref{fig:Ni_Z} we present a comparison between SiC grain data from \citet{trappitsch18} and AGB model predictions for the Ni isotopes. 
We consider AGB models including $s$-process nucleosynthesis calculated based on the Monash code \citep{karakas14b,karakas16}. The models presented here are the same models described in detail in \citet{lugaro18} and already used by those authors for comparison to Sr, Zr, and Ba in SiC grains. 
For Fe, we present similar plots as online Figure~\ref{fig:Fe_Z}. 

Single grain measurements of both Fe and Ni suffer from terrestrial contamination, which may move any data point closer to solar relatively to its true composition. For the Ni isotopes, and especially in the $\delta$(\iso{61,92,64}Ni/\iso{58}Ni) three-isotope plots, both the measured and predicted variations are significant. Furthermore, AGB predictions for different masses and metallicities lie on lines with different slopes, which are not sensitive to contamination. In contrast, for Fe only small variations are observed and predicted for  $\delta$(\iso{57}Fe/\iso{56}Fe) and the measurement uncertainties are too large to clearly distinguish the effects predicted by lines with different slopes, making it more difficult to draw conclusions based on Fe than on Ni.


For Ni, a good match is generally achieved between measurements and models both at solar and twice-solar metallicity (Figure~\ref{fig:Ni_Z}). In both cases, the most massive models (3.5 to 4 \msun) present the best match to the data. As shown in Figure~\ref{fig:Ni_PMZ} for $Z=0.03$ the full range of observations can be covered when changing the extent in mass of the mixing of protons that leads to the formation of the \iso{13}C pocket, leaving the same profile of the proton abundance \citep[see][for a detailed description of the difference between extent and profile]{buntain17} . As in the case of Sr, Zr, and Ba \citep{lugaro18}, a somewhat smaller extent in mass of the mixing provides a better coverage of the data.
However, the 4 \msun, $Z=0.014$ model should be excluded because it predicts positive $\delta$(\iso{96}Zr/\iso{94}Zr), instead of negative as seen in the grains. This is due to the opening of the branching point at \iso{95}Zr during the marginal activation of the \iso{22}Ne neutron source \citep{lugaro03b}, where in our models for the cross section of \iso{95}Zr(n,$\gamma$)\iso{96}Zr we use the relatively low value of 28 mbarn at 30 keV presented in \citet{lugaro14a}. The 3.5 \msun, $Z=0.014$ and the 3.5 and 4 \msun, $Z=0.03$ models instead provide a possible match to the measured $\delta$(\iso{96}Zr/\iso{94}Zr) values, within the nuclear uncertainties \citep[see Figure~2 of][]{lugaro18}.  

Another factor to consider are the uncertainties in the neutron-capture cross sections of the Ni isotopes. For \iso{58}Ni, \iso{62}Ni, and \iso{63}Ni, these have been recently measured at the n\_TOF facility at CERN \citep{lederer13,lederer14,zugec14}. In our temperature range of interest, from 5 to 30 keV, the values for \iso{58}Ni and \iso{62}Ni differs at most by 10\% from the values listed in the Kadonis v0.2 database \citep{dillmann06}, which we have used in the calculations, while for \iso{63}Ni the neutron-capture cross section is a factor of two higher. We tested the impact of these new rates on the models presented in Figure~\ref{fig:Ni_Z} and found variations of less than 10\% in the plotted isotopic ratios. For \iso{62}Ni we also tested the changes resulting from the uncertainties in the neutron-capture cross section and found that these changes are at most of 12\%, when varying the cross section by $2\sigma$. Therefore, no significant variations result for the models plotted in Figure~\ref{fig:Ni_Z}.

\begin{figure*}
\center{\includegraphics[scale=0.7]{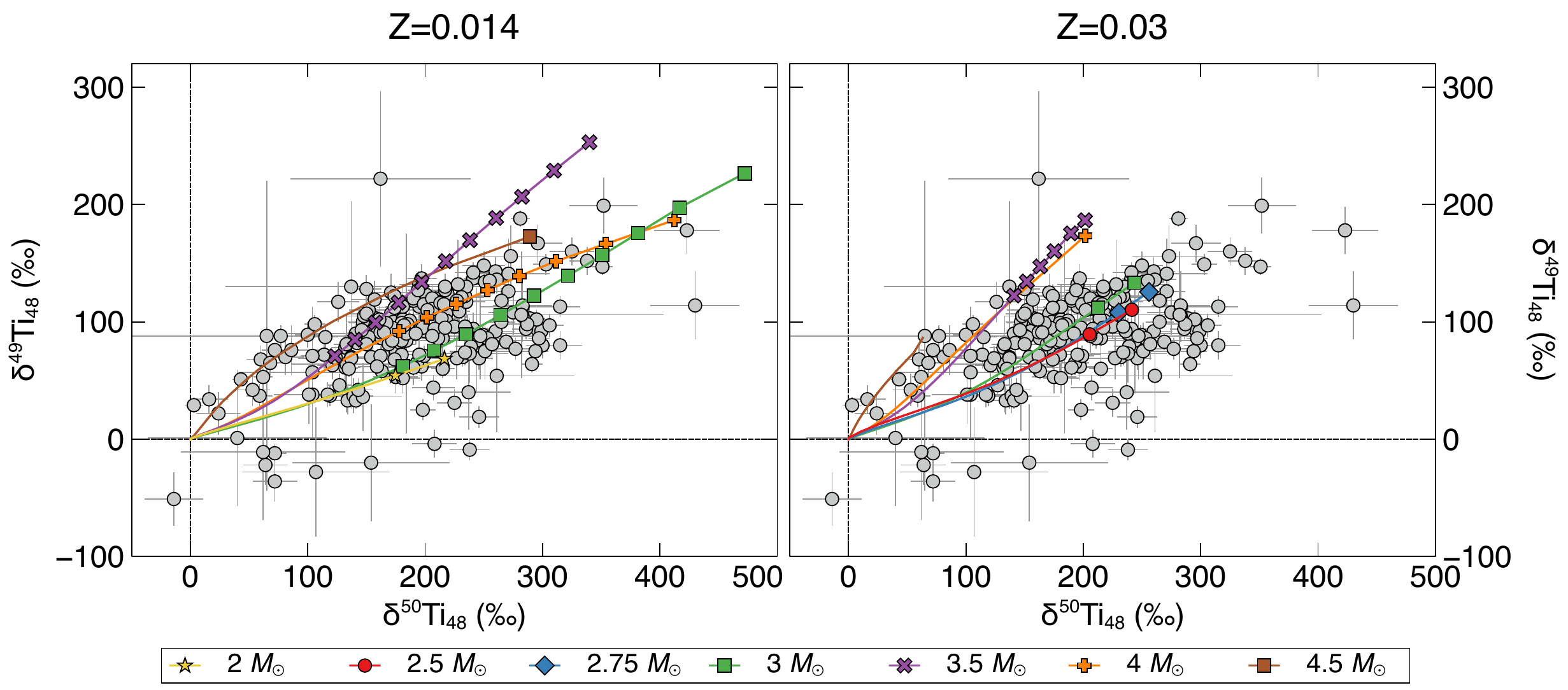}}
\caption{Same as Figure~\ref{fig:Ni_Z} but for selected Ti isotopic ratios and data for single grains of size $\gtrsim$ 2.5 $\mu$m from \citet{gyngard18}. \label{fig:Ti_Z}}
\end{figure*}

\begin{figure}
\center{\includegraphics[scale=0.7]{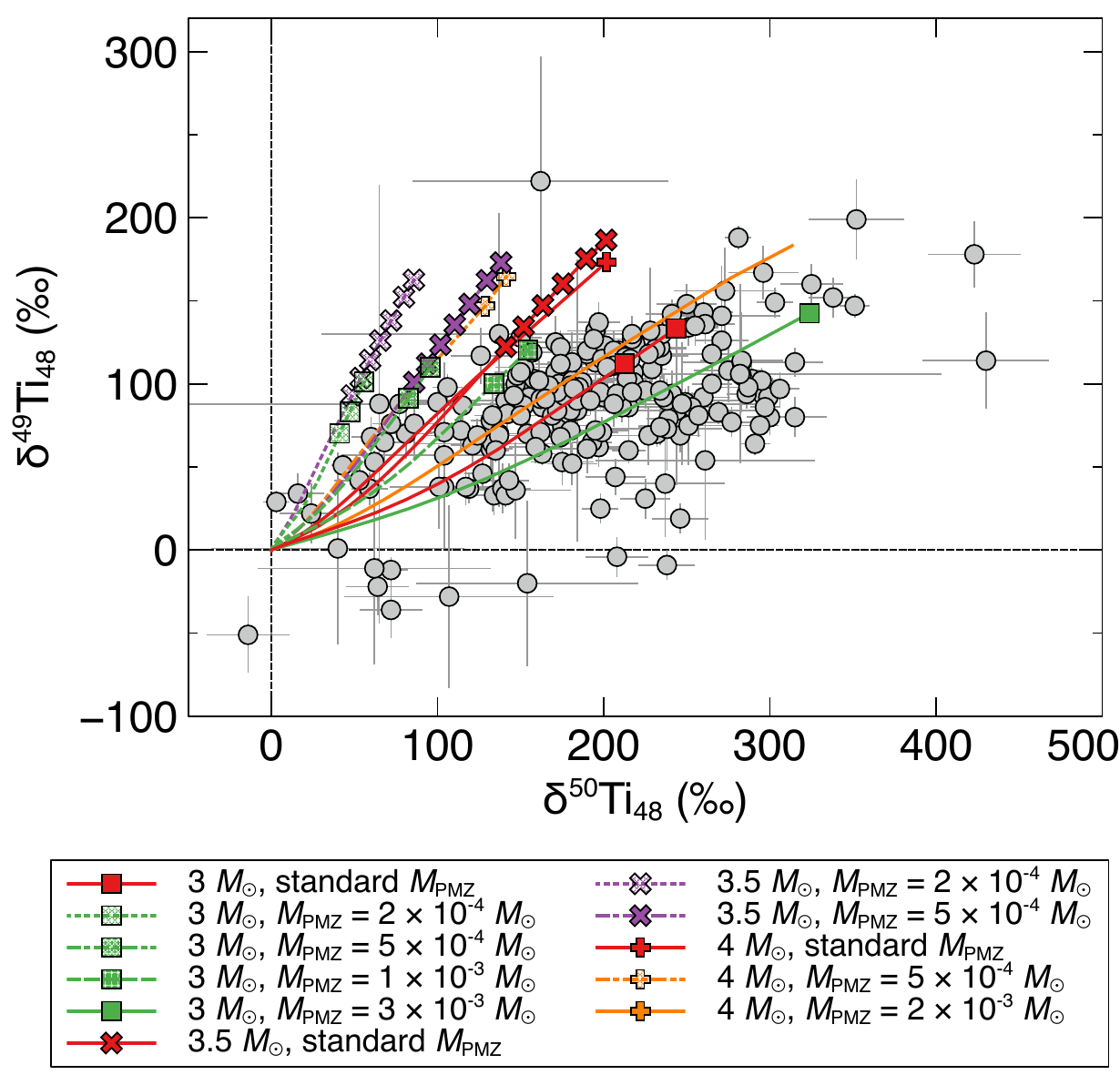}}
\caption{Same as Figure~\ref{fig:Ni_PMZ} but for selected Ti isotopic ratios and data from Figure~\ref{fig:Ti_Z}. \label{fig:Ti_PMZ}}
\end{figure}

In Figures~\ref{fig:Ti_Z} and \ref{fig:Ti_PMZ} we consider the Ti isotopic ratios that are significantly affected by AGB nucleosynthesis: $\delta$(\iso{49}Ti/\iso{48}Ti) and $\delta$(\iso{50}Ti/\iso{48}Ti). There is no significant difference in the results when changing the metallicity from solar to twice solar, although the models of higher metallicity and higher mass, which as discussed above generally provide a good match to Ni and Zr, provide a worse fit for $\delta$(\iso{49}Ti/\iso{48}Ti). The neutron-capture cross sections of \iso{49}Ti was measured in the 1970s and could suffer from strong systematic uncertainties (see the Kadonis database). When we multiplied its value by a factor of two in the 3.5 \msun\ $Z=0.03$ model we obtained a decrease of 45 in $\delta$(\iso{49}Ti/\iso{48}Ti), which results in a better agreement with the measurements. We note that both $\delta$(\iso{49}Ti/\iso{48}Ti) and $\delta$(\iso{50}Ti/\iso{48}Ti) ratios show a strong and mild, respectively, correlation with $\delta$(\iso{47}Ti/\iso{48}Ti) \citep{gyngard18}, which is not affected at all by nucleosynthesis in AGB stars and is dominated by galactic chemical evolution and heterogeneities in the interstellar medium. These effects must therefore also play a role for all of the Ti isotopic ratios in SiC grains. Furthermore, mass-independent fractionation effects have been shown to also mimic the Ti isotopic anomalies observed in the grains \citep{robert20}. Overall, AGB nucleosynthesis models cannot be used alone to compare to the Ti data. 

Finally, we note that \citet{palmerini18} also successfully matched most of the Sr, Zr, and Ba SiC data by considering a \iso{13}C pocket generated by magnetically-induced mixing, and \citet{liu18a} demonstrated that using a specific parameterization of the \iso{13}C pocket, aimed at representing the effect of such magnetically-induced mixing, also produces a match to the Ni and Sr data. Here, and in \citet{lugaro18}, we have presented another solution, obtained by keeping a basic exponential mixing profile to generate the \iso{13}C pockets, but changing the stellar metallicity to $Z=0.03$. \citet{battino19} presented yet another possible solution for the Sr data by including mixing within the \iso{13}C pocket\footnote{These authors' results do not predict enough of a deficit in \iso{96}Zr to match the data; this is related to the activation of the \iso{22}Ne neutron source in their models.}. This degeneracy of available solutions illustrates that a variety of hypotheses can be made to cover the grain data and demonstrates that it is not possible to constrain the \iso{13}C pocket, or the stellar metallicity, or the mixing processes only by comparing the grain data to the stellar models. The comparison to the Ba star spectroscopic data presented in Section~\ref{sec:translate} appears therefore to be a more reliable method to investigate the origin of the grains. However, we note that only the AGB models of 3.5-4 \msun\ and $Z=0.03$ can explain the $\delta$(\iso{92}Zr/\iso{94}Zr) values around zero and positive measured in many mainstream SiC grains without affecting the match to any other isotopic ratios \citep{lugaro18}.  

\section{Dust formation around AGB stars of super-solar metallicity}
\label{sec:dust}


\begin{figure*}
\center{\includegraphics[scale=1.]{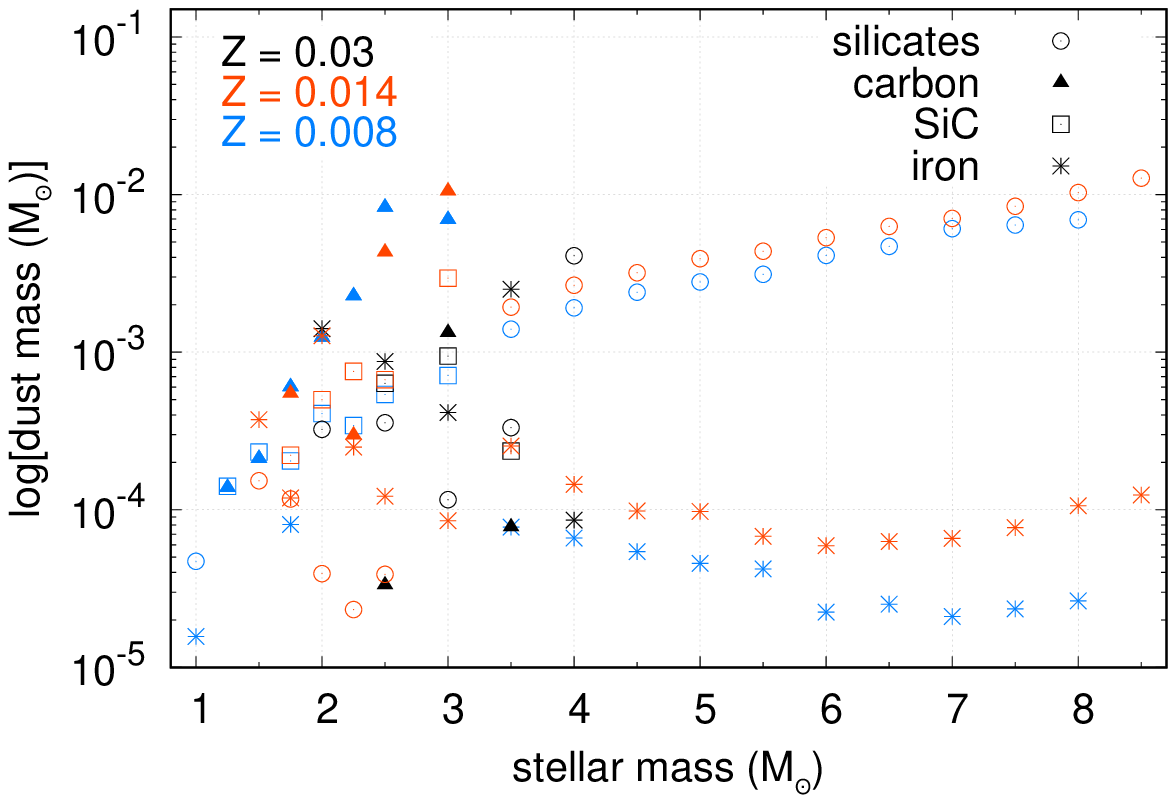}}
\caption{Amount as function of the initial stellar mass of different types of dust (silicates, carbon, SiC, and iron, different symbols) in AGB stars of metallicities from roughly half to twice solar (different colours) predicted by the ATON models. Carbon and SiC dust is produced in C-rich AGB stars in the mass range from roughly 2 to 3.5 \msun \label{fig:dust}}
\end{figure*}

Thus far, we have presented evidence that large ($\simeq \mu$m sized) mainstream SiC grains originated from AGB stars of metallicity higher than solar. Now, the question arises on why this should be the case since we may not expect a large number of low-mass stars of such metallicity to have evolved prior to the formation of the Sun. A first consideration is that the simple picture of galactic chemical evolution where metallicity increases with age is well known to be inaccurate as observations of large stellar samples show that there is no strong correlation between age and metallicity in the Galaxy. These surveys show that stars with ages between that of the Sun and roughly 8 Gyr have a spread in metallicity from 0.3 to 2.5 solar \citep{mishenina13,bensby14,hayden15}. This is currently explained by inhomogeneous GCE models, which predict a large spread in metallicity with age \citep[e.g.,][]{kobayashi11b} and/or the effect of stellar migration in the Galaxy \citep[e.g.,][]{spitoni15}. 

However, a second consideration is that the metallicity required for the parent stars of the large SiC grains is roughly twice as high as the average metallicity of stars in the solar neighbourhood. As proposed already by \citet{lewis13} in relation to Si, the shift to higher metallicity in the distribution of the parent stars of the SiC grains, relative to lower-mass, unevolved stars in the solar neighbourhood (of the same age as the higher-mass evolved parent stars of the SiC grains) may be related to a selection effect arising from the process of dust formation in AGB stars. In this section we analyse this possibility in more detail. 

We consider AGB stellar models calculated with the ATON code \citep{ventura12b}, which include the process of dust formation. For a comparison between the ATON models and the Monash models used in the previous section to discuss the $s$-process results see \citet{ventura18}. In general, depending on whether the gas is O-rich (such as in the presolar nebula and disk) or C-rich (as in some AGB stars), fundamentally different minerals can form \citep{lodders03,lodders97AIPC}. Here we are focusing on super-solar metallicity models that reach C/O$>$1, since we are interested in the formation of SiC grains at such metallicities. Other models of dust formation around AGB stars can be found, e.g., in \citet{nanni13,nanni14}, however, the AGB stars of super-solar metallicity ($Z=0.04$) presented in those papers are not C-rich. The reaching of the C/O$>$1 condition is strongly dependent on the treatment of mixing along the AGB and how it affects the efficiency of the TDU that carries C to the surface when changing the stellar metallicity. The ATON models and the Monash models discussed in Section~\ref{sec:Ni} at $Z=0.03$ both become C-rich for masses from roughly 2.5 and 4 \msun. 

Physical conditions for dust formation around AGB stars involve non-equilibrium effects with regards to both chemistry and thermodynamics \citep[including pulsations, stellar winds, possible shocks, see, e.g.,][]{sedlmayr97AIPC,hofner18}. 
The model adopted in ATON for the dust formation in AGB circumstellar envelopes was presented for the first time by \citet{gail85, gail99} and \citet{ferrarotti01,ferrarotti02,ferrarotti06}. 
In brief, the growth of dust particles is calculated on the basis of the gas density and the thermal velocity at the inner border of the condensation region during the evolution of the AGB star based on the luminosity, mass, effective temperature,  mass--loss rate, and surface chemical composition.
Dust grain formation is determined by growth and destruction rates. The first is defined by the deposition efficiency of gaseous molecules on solid seed particles (of nominal size of 1 nm) assumed here to be already formed by a prior nucleation process. For each dust species, the decomposition rate is calculated via the evaluation of the vapour pressures of the individual molecular species involved, under thermodynamic equilibrium conditions \citep{gail99}. 

In Figure~\ref{fig:dust} we show the predicted amounts of dust ejected by AGB stars of different stellar mass and metallicity for different types of dust. 
In general, the mechanism for grain growth stops when the density decreases such that gas molecules cannot interact and accrete on the existing dust anymore. However, in the case of SiC, only 55\% of the Si is available to accrete onto existing SiC dust grains because the other 45\% is already locked into the SiS molecules, which are very stable. Figure~\ref{fig:dust} shows that for C-rich AGB stars of metallicities higher than solar the amount of SiC ejected is greater or comparable to that of carbon dust. This is in contrast to the case of stars of lower metallicity, where solid carbon is instead the dominant type of dust. For example, considering the 2.5 \msun\ models, at $Z=0.03$ SiC is more than one order of magnitude greater than solid carbon, while it is 6 times lower in the case of $Z=0.014$. The amount of carbon dust depends on the number of carbon atoms available for condensation after the formation of the extremely stable CO molecules, i.e., the carbon excess with respect to oxygen. The growth of SiC particles instead is determined by the silicon abundance. At higher metallicity the C/O ratio is lower, because for the same amount of C dredged-up to the surface there is more initial O in the envelope to overcome. In these conditions, the carbon excess is smaller and SiC condensation is favoured with respect to solid carbon, as more silicon is available with respect to the lower metallicity case. As shown in Figure~\ref{fig:dust} for the C-rich AGB stars, SiC is the dominant species only if $Z=0.03$ , for the models of 2.5 and 3.5 \msun. For the 
3.0 \msun\ model the amount of SiC is comparable to that of solid carbon. Relative to the $Z=0.014$ models, the 2.5 \msun\ case produces the same amount of SiC, while the 3 \msun\ model produced 3 times less. (The 3.5 \msun\ models remain O rich and does not produce significant C-rich dust.)
This is in disagreement with the distribution shown in the bottom panel of Figure~\ref{fig:Bastars}, which indicates that AGB stars at $Z=0.03$ should produce (at least $\sim$50\%) more SiC relative to the case at $Z=0.014$.

It should be noted, however, that dust formation models in general carry some uncertainties, in particular related to the mass loss. In the model discussed here the dust formation rate increases with the mass loss rate because the mass loss affects the density of the wind (via mass conservation), and thus the number of gaseous molecules available to condense into dust. Carbon stars lose mass at higher rates than O-rich AGB stars because once the surface carbon exceeds oxygen the surface molecular opacities become extremely large \citep[see, e.g.,][]{ventura10}.
These opacities favour the expansion of the external regions, which become less and less gravitationally bound to the central star, thus increasing the rate of mass loss and consequently of dust production. The increase in the molecular opacities does not depend linearly on the C/O ratio and even a small difference (with 3 \msun\ typical values, from 1.1 in the $Z=0.03$ model to 1.25 in the $Z=0.014$ model) is sufficient to provoke a significant difference in the rate of mass loss experienced and in turn of the total dust production: the 3 \msun\ $Z=0.03$ and $Z=0.014$ models, for example, produce a total dust mass of $2.82 \times 10^{-3}$ and $1.35 \times 10^{-2}$ \msun, respectively, a factor of five difference.
To illustrate the significant effects of these uncertainties we consider for comparison the models presented by \citet{ferrarotti06}, which  
are not based on the use of C/O dependent molecular opacities. In this case, the most relevant quantity for the formation of SiC is the silicon abundance. Predictions from these models are therefore different from those presented here, for example, the 3 \msun, $Z=0.04$ models of \citet{ferrarotti06} produces four times more carbon dust than SiC and, relative to the $Z=0.02$, a similar amount of total dust and four times more SiC. 

The other crucial feature, discussed already in Sec.~\ref{sec:size} is the size of the dust grains. In the ATON models, all SiC grains reach a maximum diameter of 0.26\, $\mu$m, with an approximate lower limit of $0.16\,\mu$m. It is not possible to form larger grains because all the gaseous silicon is already either locked into SiS molecules or condensed into SiC grains. These maximum dimensions correspond to grains within the meteoritic KJA fraction and exclude the large grains belonging to the KJE and KJG fractions that we are considering here to originate in stars of super-solar metallicity. While we found that the velocity with which the gas enters the condensation region does not significantly influence the grain size, the density of the seed dust grains, assumed in the model to be already present in the AGB envelope and to act as seeds onto which larger dust can grow, has a strong impact. Because the amount of available Si to condense into SiC is fixed, necessarily, if there are fewer seeds the final SiC grain size is larger. 
In the models presented above we used a value for number of seed dust grains of $N_{\rm d} = 10^{-13} N_{\rm H}$, where $N_{\rm H}$ is the number of H atoms. This number reflects, as an order of magnitude, the average estimate based on the analysis of a sample of Galactic giants presented by \citet{knapp85} and is commonly assumed in the models to be independent of the dust species and of the stellar metallicity, which is not necessarily correct. If we decrease $N_{\rm d}$ by a factor of 100 or 1000 the grains reach a size of 1.2 and 4 $\mu$m, respectively, which would cover the meteoritic SiC of fractions KJD--KJE and KJG, respectively (Table~\ref{tab:size}). Therefore, if the number of seeds decreases with increasing the metallicity, this would results in a selection effect where the larger SiC grains preferentially form in stars of higher metallicity.

\section{Summary, discussion, and conclusions} 

We have presented a new approach to identify the origin of meteoritic stardust mainstream SiC grains from C-rich AGB stars based mostly on spectroscopic observations of $s$-process enriched Ba stars. This approach allows us to reach more robust conclusions because models of nucleosynthesis in AGB stars are prone to many uncertainties, which are bypassed with our new method. For Ba stars we selected [Ce/Y] as the representative signature of the $s$-process nucleosynthesis experienced by their binary AGB companions. For the SiC grains we selected $\delta$(\iso{88}Sr/\iso{86}Sr) as the representative signature, since it involves the isotopic ratio affected by the number of neutron captured per Fe seed that shows the largest range of variations in the grain data.  

Our main results are the following:

\begin{enumerate}

\item{Due to the existence of nuclei with magic number of neutrons on the $s$-process path, the $s$ process necessarily produces negative [Ce/Y] ratios when  $\delta$(\iso{88}Sr/\iso{86}Sr) is also negative. Ba stars (and $s$-process AGB stellar models) show a clear trend of [Ce/Y] decreasing as the metallicity increases \citep{cseh18}, and it is statistically more likely for Ba stars of metallicity higher than solar to show negative [Ce/Y] ratios (Figure~\ref{fig:Bastars}). SiC grains show a range of $\delta$(\iso{88}Sr/\iso{86}Sr) decreasing with increasing the grain size and down to negative for grains of size $\gtrsim \mu$m. Therefore, the larger grains should have originated from AGB stars of higher metallicity than the smaller grains.} 

\item{The isotopic compositions of Ni, Sr, Zr, and Ba in $\mu$m-sized grains is well matched by AGB models of metallicity higher than solar \citep[see also][]{lugaro18}. The composition of Si is mostly affected by the chemical evolution of the Galaxy and also points to the grains coming from stars of metallicity higher than solar \citep[see also][]{lewis13}, in particular as their size increases \citep{amari00}. For Ti, galactic chemical evolution as well as mass-dependent fractionation effects \citep{robert20} will need to be considered together with AGB models.}

\item{In AGB stellar models of metallicity higher than solar, SiC is typically the dominant type of dust produced, however, these models produce lower absolute masses of SiC, and of dust in general, than the models of solar metallicity. In general, the AGB models of dust formation do not predict the grain sizes that are observed for the large SiC grains, unless the number of dust seeds is decreased from $10^{-13}$ to $10^{-15,-16}$ of the number of H atoms.}

\end{enumerate}

Since the most likely sites of the origin of large ($\mu$m-sized) mainstream SiC stardust grains are AGB stars of metallicity higher than solar (1), and that to produce grains of such size we need to decrease the number of dust seeds (3), we conclude with the hypothesis that the number of dust seeds in AGB stars should decrease when increasing the metallicity, and that larger SiC dust grains should be present in AGB stars of metallicity higher than solar. Testing this hypothesis with observations of AGB stars using the millimeter/submillimeter telescope ALMA or the mid-Infrared MATISSE may be possible by observing nearby AGB stars; however, this is not trivial. First, it is difficult to establish the metallicity of AGB stars. One way to indirectly sample variations with metallicity would be to investigate objects at different distance from the galactic centre, given that a metallicity gradient exists with the galactic longitude. For example, \citet{groenewegen02} found a decrease of the expansion velocity with galactic distance in a sample of 330 infrared carbon stars, possibly due to a higher gas-to-dust ratio outside the solar circle than inside, and related to the metallicity gradient. Second, it is difficult to firmly establish grain sizes from observations because not only the size but also radiative transfer effects due to the detailed density stratification of dust and the opacity profile of SiC may change the shape of the SiC spectral feature. Finally, it is difficult to disentangle the properties of SiC from those of amorphous carbon dust. Using interferometric imaging it may be possible to see a change in the wavelength dependent appearance, if SiC really is located further in than carbon \citep[as observed for example for the dust mineralogy around young stellar objects,][]{vanboekel06}. The number of SiC seeds could be also derived by comparing the depletion of molecular SiC between targets of similar mass loss. However, SiC molecular features are not obvious and in general their observations involves a large number of uncertainties. Grain sizes can be estimated from polarimetry (combined with interferometry) as shown for O-rich grains \citep{norris12}. Also this, however, involves a large number of parameters and uncertainties. In parallel, more work is needed from the theoretical point of view, both different dust formation models for super-solar metallicities \citep[e.g., extending the models of][]{bladh19}, and the consequences on the mass loss and spectral energy distribution of larger and more abundant SiC grains in stars of higher metallicity need to be considered. 

One important point to keep in mind is the definition of metallicity in relation to all the different aspects we are considering here, and in relation to the solar metallicity. On the one hand, for Ba stars metallicity refers to [Fe/H] and since the abundance of Fe is relatively well established in the Solar System, we can confidently conclude that Ba stars with an Fe abundance twice the solar abundance show predominantly negative [Ce/Y]. On the other hand, when we consider stellar models, C and O are the main elements contributing to the metallicity. The abundances of these elements strongly affect the opacity, both deep in the star where the $s$-process occurs and in the envelope, where the dust forms. Furthermore, since the lower solar O abundance determined via spectroscopy is still in disagreement with helioseismology data \citep[e.g.,][]{vinyoles17}, the overall solar metallicity is still uncertain. Thus, whether the $Z=0.03$ AGB models represent stars of twice
solar metallicity based on the lower O abundance from \citet{asplund09} (giving $Z=0.014$) or 1.5 times solar based on the O abundance from the older compilation by \citet{anders89} ($Z=0.02$) is uncertain. As a result, we do not consider it a strong inconsistency that the [Ce/Y] ratios in Ba stars indicate that most large grains should come from stars with Fe twice solar, while \citet{lewis13}, based on models of the galactic chemical evolution of Si, derived that the majority of SiC grains should have formed in AGB stars of metallicity from solar to 70\% above solar. We note that \citet{lewis13} did not make any distinction among grains of different sizes, while this point should be considered in the future. Similar studies should also be performed in relation to silicate grains that originated from O-rich AGB stars and show a similar Si isotopic distribution as SiC \citep{hoppe18}, although no $s$-process isotopic ratios are available for such grains. More detailed studies are also needed to verify the compatibility of our results with the composition of the noble gases, and of other elements such as Mo, W, Hf, and Pb. 

Based on the fact that bulk meteorite analyses show smaller magnitude $s$-process variations in the heavier (second peak) relatively to the lighter (first peak) refractory $s$-process elements, \citet{ek19} suggested that high-metallicity AGB stars may have been a dominant source of stardust in the early Solar System. This suggestion was based on the fact that AGB stars of higher metallicity produce less second peak $s$-process elements, relatively to the first peak (see Sec.~\ref{sec:bastars}). If this idea is correct, our results further indicate that the large SiC grains represent the material from high-metallicity AGB stars that is needed to interpret the bulk rock anomalies. In this case, these large grains should  represent a significant fraction of presolar stardust in the Solar System. Large grains obviously carry more material than small grains, but their impact will also depends on numbers. By number, the amount of grains of different sizes appears to follow a power-law distribution above $\sim$0.6 $\mu$m \citep[see Figure 8 of][]{amari94}, although this result is not conclusive because many smaller grains could have been lost because of the chemistry used to make the K-series residues. A more detailed analysis is required. Theoretically, if large SiC grains survive longer in the interstellar medium than small SiC grains, then they may have been preferentially present in the presolar cloud. More detailed laboratory and theoretical investigations on this topic are required.  Although challenging,  more data on Sr and Ba isotopes in different Solar System materials will advance our understanding of the presence and evolution of presolar dust and stardust in the solar proto-planetary disk.
 

\acknowledgments
We thank Jacqueline Den Hartogh, Andr\'es Yag\"ue and M\'aria Peto for discussion, Nan Liu for sharing unpublished data, and the referee for the careful reading and commenting the paper, which helped us to improve it. M.L. and B.Cs. acknowledge the financial support of the Hungarian National Research, Development and Innovation Office (NKFI), grant KH\_18 130405. This work is also supported by the Lend\"ulet grant (LP17-2014) of the Hungarian Academy of Sciences, grants GINOP-2.3.3-15-2016-00003, K-119517, and K-115709 of the NKFI, the City of Szombathely under agreement No. S-11-1027, and the European Research Council ERC-2015-STG Nr. 677497. A.I.K. acknowledges financial support from the Australian Research Council (DP170100521). We thank the ChETEC COST Action (CA16117), supported by COST (European Cooperation in Science and Technology). Part of this work was performed under the auspices of the US Department of Energy by Lawrence Livermore National Laboratory under Contract DE-AC52-07NA27344 and was supported by the LLNL-LDRD Program under Project 20-ERD-030. LLNL-JRNL-802744. M.P. acknowledges significant support to NuGrid from NSF grant PHY-1430152 (JINA Center for the Evolution of the Elements) and STFC (through the University of Hull's Consolidated Grant ST/R000840/1), and access to {\sc viper}, the University of Hull High Performance Computing Facility. Parts of this research were supported by the Australian Research Council Centre of Excellence for All Sky Astrophysics in 3 Dimensions (ASTRO 3D), through project number CE170100013 .


\vspace{5mm}

\bibliographystyle{yahapj}

\begin{figure*}
\center{\includegraphics[scale=0.7]{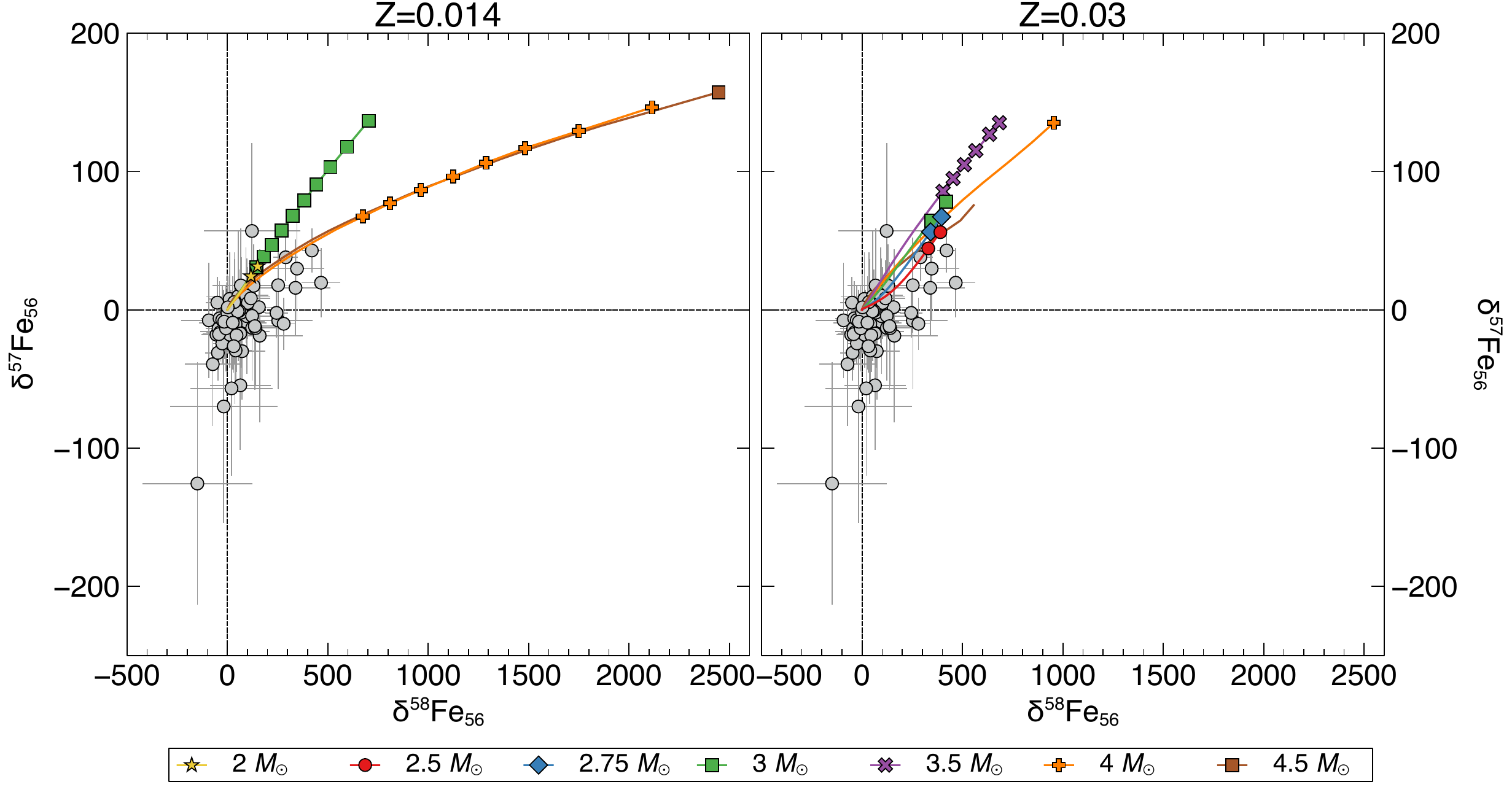}}
\caption{Same as Figure~\ref{fig:Ni_Z} but for selected Fe isotopic ratios. \label{fig:Fe_Z}}
\end{figure*}





\end{document}